\documentclass[onecolumn,floats,prb,aps,showpacs]{revtex4}
\usepackage{graphicx}
\usepackage{amsfonts}
\usepackage{amsmath}
\usepackage{amssymb}
\usepackage{exscale}
\usepackage{color}

\begin{document}

\title{Thermal suppression of surface barrier in ultrasmall superconducting
structures}
\author{W. V. Pogosov}
\affiliation{Institute for Theoretical and Applied Electrodynamics, Russian Academy of
Sciences, Izhorskaya 13, 125412 Moscow, Russia}
\date{\today }

\begin{abstract}
In the recent experiment by Cren \textit{et al.} [Phys. Rev. Lett. 
\textbf{102}, 127005 (2009)], no hysteresis for vortex penetration and
expulsion from the nano-island of Pb was observed. In the present paper, we
argue that this effect can be associated with the thermoactivated
surmounting of the surface barrier by a vortex. The typical entrance (exit)
time is found analytically from the Fokker-Planck equation, written in the
form suitable for the extreme vortex confinement. We show that this time is
several orders of magnitude smaller than 1 second under the conditions of
the experiment considered. Our results thus demonstrate a possibility for
the thermal suppression of the surface barrier in nanosized low-$T_{c}$
superconductors. We also briefly discuss other recent experiments on
vortices in related structures.
\end{abstract}

\maketitle

\section{Introduction}

Vortex nucleation in an infinite type-II superconductor becomes
energetically favorable at lower critical field $H_{c1}$. It is however
clear that real superconductors are always bounded, so that vortices can
appear in a sample, at low temperatures, only by penetration through
surfaces. This process was for the first time considered by Bean and
Livingston\cite{BL} within the framework of the London approximation.
In their approach, single vortex was represented by a straight line entering
a semi-infinite sample through its surface. Bean and Livingston have found a
surface barrier preventing such an entry. The barrier disappears when $H\ $\
is close to the thermodynamical critical field $H_{c}$, which is much larger
than $H_{c1}$ in superconductors with high values of the Ginzburg-Landau
parameter $\kappa $ ($H_{c}/H_{c1}\sim \kappa /\ln \kappa $). Physically,
Bean-Livingston barrier is a consequence of the competition between the two
factors: (i) the interaction between the vortex and the Meissner current,
which tries to push the vortex line inside the sample; (ii) the Magnus
attraction of a vortex to the surface, which can be understood in terms of
the interaction between the vortex and its mirror image ("antivortex").
Similarly, surface barrier exists for the vortex exit from the sample.
Bean-Livingston barrier plays an important role in the vortex dynamics near
the superconductor surface and leads to pronounced hysteresis effects\cite%
{Brandt}.

As any potential barrier, Bean-Livingston barrier can be surmounted by
thermal activation. This phenomenon was addressed by Petukhov and Chechetkin%
\cite{PetChech}, who have considered vortex nucleus as a half-loop
spreading from the surface into the bulk. The profile of the surface barrier
was calculated within the London approximation and the typical penetration
time was estimated by using Fokker-Planck equation. The resulting formula is
of Arrhenius type with the height of the surface barrier entering through
the exponent. After substituting parameters typical for conventional low-$%
T_{c}$ materials, this height was estimated to be as large as $10^{3}$-$%
10^{5}$ in terms of $k_{B}T$, from which it was concluded that thermally
activated vortex penetration is practically unobservable.

However, two decades later, after the discovery of high-$T_{c}$ materials,
thermal activation over the surface barrier has attracted great attention.
It was shown\cite{Kopylov} by Kopylov \textit{et al.} that thermal
surmounting of the barrier becomes possible in layered high-$T_{c}$
superconductors, where the mixed state structure is represented by quasi-2D
pancake vortices. Burlachkov\cite{Burlachkov} has analyzed the same
phenomenon for high-$T_{c}$ superconductors of YBaCuO type with 3D vortex
lines. He has demonstrated that the height of the surface barrier in these
compounds is dramatically suppressed, compared to low-$T_{c}$ materials, to $%
10-100k_{B}T$, which leads to experimentally observable rates for thermally
activated vortex penetration and expulsion. Indeed, these conclusions agree
with numerous experiments, see e.g. Ref \cite{exper}, as well as the
theoretical paper \cite{TheorAll}.

Thus, it is generally believed that thermally activated flux penetration and
expulsion in low-$T_{c}$ superconductors is not possible, while for the case
of high-$T_{c}$ superconductors it is quite standard. However, in the very
recent experiment\cite{Cren} by Cren \textit{et al.} no hysteresis for
vortex entry and exit was found in a superconducting nano-island of Pb
having an extremely small thickness ($\approx 5.5$ nm). Lateral dimension of
this sample was so small ($\sim 100$ nm) that it could accommodate only one
vortex before turning to the normal state. This regime was referred in Ref. %
\cite{Cren} as ultimate vortex confinement.

The aim of the present paper is to show that thermal fluctuations can be
responsible for the effective suppression of the surface barrier observed in
the experiment of Cren\cite{Cren} \textit{et al.}. This cannot be
firmly proved by using previously developed approaches to the thermal
activation over the surface barrier, since all these schemes are based on
the London theory, while this theory is not applicable for small-sized
samples with lateral dimensions of the order of the vortex core size.
Moreover, under these conditions, vortex position is not an appropriate
variable, because an initial stage of vortex nucleation is associated with a
"nascent" vortex, considered for the first time by Kramer\cite{Kramer}%
. "Nascent" vortex represents an area at the sample's edge, where the order
parameter is suppressed, but it does not vanish yet. As we show, the height
of the potential barrier, under the experimental conditions of Ref. %
\cite{Cren}, is mostly due to the vortex core nucleation rather than
because of the motion of already formed vortex towards the island center.
Thus, the origin of the surface barrier in the limit of a very small sample
size is not identical to the one for the traditional Bean-Livingston
barrier. This feature has already been discussed in Ref. \cite{Doria}.

In the present paper, in order to circumvent these difficulties, we use the
following alternative scheme. (i) We first solve Ginzburg-Landau equations
approximately by using lowest Landau level approximation for the order
parameter and determine the surface barrier in terms of the populations of
corresponding levels. The island we consider is so small that it can
accommodate only one vortex. Hence, to describe a barrier with a reasonable
accuracy, it is sufficient to take into account only two levels (with
winding numbers 0 and 1), which correspond to the vortex-free and one-vortex
states. Populations of these levels are then used as "good" variables
instead of the vortex position. Notice that there are no essential
difficulties in extending the number of variables by taking into account
more levels, which can be necessary when considering larger structures. (ii)
We then estimate "viscosity" coefficients associated with the motion of
order parameter projected on the two Landau levels. Demagnetization effects
are included into the consideration. (iii) At the final step, we solve the
backward Fokker-Planck equation, again in terms of the Landau levels
populations. Our main result is an analytical expression for the typical
time for thermally activated vortex entry (exit). Notice that the lowest
Landau level representation for the order parameter in mesoscopic
superconductors was proposed in Refs. \cite{VVM,SchPee,Palacios} and then
used very widely.

Our calculations show that the height of the surface barrier, under the
conditions of the experiment considered, is approximately\ $20k_{B}T$ that
is more typical for high-$T_{c}$ materials than for conventional low-$T_{c}$
superconductors. The major reasons for the potential barrier to be not so
high are (i) very small island thickness, (ii) very small island lateral
dimensions which result in the substantial suppression of the order
parameter, (iii) relatively high sample's temperature\cite{Cren} ($%
\approx 4.3$ K). The preexponential "attempt time" is quite small \ - mainly
because of the small system sizes, which result in a weak magnetic response
of the island. This time was estimated to be of the order of 10$^{-15}-$10$%
^{-14}$ s which is much smaller than the similar quantity\cite{Kopylov}
for high-$T_{c}$ layered superconductors with pancake vortices ($\sim
10^{-12}$ s). The resulting first passage time $\tau $ both for the vortex
entry and exit is also quite small, $\tau \sim $10$^{-5}-$10$^{-4}$ s.

Thus, the results we here present demonstrate that the thermally-assisted
suppression of the surface barrier can occur in nanosized samples made of 
\textit{low-}$T_{c}$\textit{\ superconductors}, which are already available
for experimental investigations.

Since nanosized superconductors might be perspective for technological
application (see e. g. Ref \cite{Shifter,Mishko}) and are even already
used (for instance, for single-photon detection purposes\cite{Bartolf}%
), the control of fluctuations in such structures is quite important.

This paper is organized as follows. In Section II, we study surface barrier
within the Ginzburg-Landau theory. In Section III we estimate viscosity
coefficients, with taking into account demagnetization effects. In Section
IV, we solve the backward Fokker-Planck equation and determine first passage
times both for the vortex penetration and expulsion. We conclude in Section
V.

\section{Surface barrier}

In this section, we determine approximately the height of the potential
barrier for vortex exit and entry under the conditions of experiment\cite%
{Cren}. Our results are applicable as long as the sample's radius remains
comparable to the coherence length $\xi (T)$.

\subsection{Main parameters}

The superconducting island studied in Ref. \cite{Cren} had a
"smoothed" hexagonal shape with a small thinner region in its center, as
seen from Fig. 1(b) of Ref. \cite{Cren}. The smallest lateral size of
the sample was 110 nm, while the thickness was $d=5.5$ nm. Sample
temperature was $T=4.3$ K. The transition between the vortex-free and
one-vortex states, as obtained by STS measurements, occurred at the applied
field of $0.235$ T. The transition was reversible within 3 \%. The
reversibility implies that at this field energies of vortex-free and
one-vortex phases become equal.

There was a disordered layer between the island and the Si substrate, which
limitates\cite{Cren} the quasiparticle mean free part to $l\approx 2d$%
. The zero-temperature coherence length was estimated\cite{Cren} by
using an expression for dirty superconductors $\approx \sqrt{\xi _{0}l}$,
where $\xi _{0}=80$~nm is the coherence length in bulk Pb. This yields $%
\approx 30$ nm. The coherence length $\xi (T=4.3K)$ should be larger, but it
is not easy to find its precise value theoretically because of the small
system sizes (and nontrivial shape), which makes the microscopic physics
quite complex\cite{Tinkham}. However, it was possible to extract the
coherence length directly from the measured zero-bias conductance profile in
the vortex core, which yielded the value of $40-45$ nm \cite{Dima}. In
this paper, we assume a value of $48$ nm. For the penetration depth, we use
the usual expression\cite{Tinkham} for dirty superconductors, $\lambda
(T)\simeq 0.615\lambda _{0}\sqrt{\frac{\xi _{0}/l}{(1-T/T_{c})}}$ where $%
\lambda _{0}\simeq 40$~nm is the penetration depth in bulk Pb, $T_{c}=7.2$
K. This gives $\lambda (4.3K)\simeq 102$ nm. Notice that the use of the
prefactor $0.615$ is, in fact, rather relative. The Ginzburg-Landau
parameter, corresponding to the above numbers, is $\kappa \approx $ $2$.
When considering the diamagnetic response of the island, one has to keep in
mind that effective $\kappa $ is much higher\cite{Pearl}, since $d\ll
\lambda (T)$.

Although the island has a rounded hexagonal shape, it is actually quite
similar to a disc (see Fig. 1(b) of Ref. \cite{Cren}). Therefore, we
model it by the disc. As for its diameter, we use the value of 150 nm, which
corresponds approximately to the largest ("bottom") lateral dimension of the
island. This value together with $48$ nm for the coherence length will allow
us to reproduce the experimentally found field for the transition between
the vortex-free and single-vortex states ($0.235$ T).

Note that the radius of the disc in terms of the coherence length is
approximately 1.55. This regime indeed corresponds to the "ultimate vortex
confinement": the disc is able to accommodate only one vortex before the
transition to the normal state, see e.g. Ref. \cite{Pogosov}.

\subsection{Basic estimates}

Let us make some basic estimates. We compare the thermal energy $k_{B}T$
with the energy required to create a vortex nucleus, i.e., to suppress the
order parameter to zero within the volume $\sim \xi (T)^{2}d$. The similar
idea is applied when calculating the Ginzburg-Levanyuk number%
\cite{Varlamov}. Within the Ginzburg-Landau theory, the energy gain
due to the vortex nucleus is 
\begin{equation}
\sim \frac{H_{c}(T)^{2}}{2\mu _{0}}\xi (T)^{2}d,  \label{1}
\end{equation}%
where $H_{c}(T)$ is the thermodynamical critical field, given by%
\begin{equation}
H_{c}(T)=\frac{\Phi _{0}}{2\pi \sqrt{2}\xi (T)\lambda (T)}.  \label{2}
\end{equation}%
For the experiment performed at $T=4.3K$ we obtain the following estimate
for the ratio: 
\begin{equation}
k_{B}T\left( \frac{H_{c}(T)^{2}}{2\mu _{0}}\xi (T)^{2}d\right) ^{-1}\sim
10^{-2},  \label{3}
\end{equation}%
which is basically not so small. Notice that the order parameter is strongly
suppressed in the vicinity of the transition between vortex-free and
one-vortex states. This should reduce condensation energy and enhance the
ratio (3) in several times. Also, we expect that the barrier height to be
much lower than the above estimate due to an additional reason: vortex
nucleation is associated mainly with the \textit{redistribution} of the
density of superconducting electrons inside the island rather than with the
change of their total number. Thus, the ratio of $k_{B}T$ and the barrier
height has to be of the same order as in high-$T_{c}$ superconductors\cite%
{Burlachkov}, where rates for thermally activated flux penetration and
expulsion are observable. Below we show that this is indeed the case.

\subsection{Calculation of the surface barrier}

We start with the dimensionless Ginzburg-Landau functional for the energy of
the disc in the superconducting phase compared to the normal state:

\begin{equation}
F=\frac{H_{c}(T)^{2}}{\mu _{0}}\xi (T)^{2}d\int_{0}^{2\pi }d\varphi
\int_{0}^{r_{0}}rdr\left( -\left\vert f\right\vert ^{2}+\frac{1}{2}%
\left\vert f\right\vert ^{4}+\left\vert \left( -i\mathbf{\nabla }-\mathbf{a}%
\right) f\right\vert ^{2}\right) ,  \label{4}
\end{equation}
where integration is performed over the disc cross section (cylindrical
coordinate system is used), $r_{0}=R/\xi (T)$ is the dimensionless radius of
the disc, $f$ is the dimensionless order parameter, $a=h_{e}r/2$ is the
vector potential, which has only an azimuthal component, $h_{e}=H/H_{c2}(T)$
is the external magnetic field measured in terms of the upper critical field 
$H_{c2}(T)=\Phi _{0}/2\pi \xi (T)^{2}$, and all the distances are measured
in units of $\xi (T)$. Due to small disc sizes, the external magnetic field
penetrates the sample almost totally, so that the magnetic field inside the
disc is essentially equal to the external field. Since the disc thickness $d$
is much smaller than $\xi (T=4.3K)$, the order parameter does not vary in $z$
direction. Hence, the problem both for the order parameter and barrier
height is effectively two-dimensional.

The order parameter can be represented as a Furrier expansion 
\begin{equation}
f(r,\varphi )=\sum_{n}c_{n}\varphi _{n}(r)e^{-n\varphi }.  \label{5}
\end{equation}%
In the vortex-free state, only one coefficient among $c_{n}$'s is nonzero,
namely $c_{0}$. In the one-vortex state, a nonzero coefficient is $c_{1}$.
The disc we consider is so small (in terms of $\xi (4.3K)$) that it can
accommodate only one vortex. Therefore, with good accuracy, we can map the
order parameter on the subspace with only two nonzero components:%
\begin{equation}
f(r,\varphi )\simeq c_{0}\varphi _{0}(r)+c_{1}\varphi _{1}(r)e^{-i\varphi }.
\label{6}
\end{equation}%
The superposition (6) actually describes vortex exit or entrance. If we
start with the vortex-free state ($c_{1}=0$), an increase of $c_{1}$ first
leads to the suppression of the order parameter at some spot on the disc
edge. This suppression can be interpreted as a formation of a vortex nucleus
or a "nascent" vortex \cite{Kramer}. At certain value of $c_{1}$, the
order parameter in the center of this nucleus gets suppressed to zero, and
the phase of the order parameter now changes by $2\pi $ when turning around
this topological defect. Thus, the real vortex is formed. Further increasing
of $c_{1}$ leads to the vortex displacement towards the disc center, with
its position $r_{v}$ given simply by the condition: $\left\vert
c_{0}\right\vert \varphi _{0}(r_{v})-\left\vert c_{1}\right\vert \varphi
_{1}(r_{v})$, which follows directly from Eq. (6). If we now start with the
one-vortex state ($c_{0}=0$), an increase of $c_{0}$ results in the vortex
displacement towards the disc edge until it leaves the island.

Further simplifications come by using, for $\varphi _{0}(r)$ and $\varphi
_{1}(r)$, the eigen functions for the kinetic energy operator corresponding
to winding numbers $L=0$ and $L=1$, respectively, and to zero radial quantum
number (nodeless wave functions). The latter corresponds to the lowest eigen
value for the kinetic energy operator, see e.g. Ref. \cite{Yampol}.
This choise for $\varphi _{0}(r)$ and $\varphi _{1}(r)$ can be justified by
noting that at the transition to the normal state, the first Ginzburg-Landau
equation can be linearized, i.e., reduced to the eigen value equation for
the kinetic energy operator (with the eigen value 1). Therefore, in the
vicinity of the transition, $\varphi _{n}(r)$ can be closely approximated by
the the eigen functions for the same operator. This regime is indeed
realized in the superconducting disc under conditions of "ultimate vortex
confinement", where the transition between the one-vortex and vortex-free
states occurs at high fields, $H\approx H_{c2}(T)$, so that the density of
superconducting electrons is strongly suppressed. In cylindrical
coordinates, eigen value equation reads%
\begin{equation}
-\frac{1}{r}\frac{d}{dr}\left( r\frac{d\varphi _{L}}{dr}\right) +\varphi
_{L}\left( \frac{h_{e}r}{2}-\frac{L}{r}\right) ^{2}=\varepsilon _{L}\varphi
_{L}.  \label{7}
\end{equation}%
This equation can be solved analytically:%
\begin{equation}
\varphi _{L}=r^{L}\varepsilon _{L}^{L/2}\exp \left( -\frac{h_{e}r^{2}}{4}%
\right) \Phi \left( \frac{h_{e}-\varepsilon _{L}}{2h_{e}},L+1,\frac{%
h_{e}r^{2}}{2}\right) ,  \label{8}
\end{equation}%
where $\Phi $ is Kummer function. Eq. (7) has to be supplemented by the
boundary condition at the superconductor/vacuum interface%
\begin{equation}
\left. \frac{d\varphi _{L}}{dr}\right\vert _{r=r_{0}}=0.  \label{9}
\end{equation}%
Values of $\varepsilon _{L}$ for given $h_{e}$, $r_{0}$, and $L$ can be
found numerically from Eqs. (8) and (9).

We now treat $c_{0}$ and $c_{1}$\ as variational parameters and calculate
disc energy, as given by Eq. (4), as a function of these two parameters.
Thus the energy of the system is obtained as a biquadratic function of
variational parameters $c_{0}$ and $c_{1}$. The resulting expression can be
minimized with respect to $c_{0}$ and $c_{1}$. Without loss of generality, $%
c_{0}$\ and $c_{1}$\ can be taken real and positive. Although this method is
not exact, we expect that it allows one to obtain rather accurate estimate
of the potential barrier height for such a small island with the radius only
slightly exceeding the vortex core size \cite{Peeters-barrier}.

The energy of the island, obtained from Eqs. (4)-(9), is given by%
\begin{equation}
F=\frac{H_{c}(T)^{2}}{\mu _{0}}\xi (T)^{2}d\left( c_{0}^{2}\alpha
_{0}+c_{1}^{2}\alpha _{1}+c_{0}^{4}\beta _{0}+c_{1}^{4}\beta _{1}+2\gamma
c_{0}^{2}c_{1}^{2}\right) ,  \label{10}
\end{equation}%
where ($L=0$, $1$)%
\begin{eqnarray}
\alpha _{L} &=&(\varepsilon _{L}-1)\int_{0}^{r_{0}}\varphi _{L}^{2}rdr,
\label{11} \\
\beta _{L} &=&\frac{1}{2}\int_{0}^{r_{0}}\varphi _{L}^{4}rdr,  \label{12} \\
\gamma &=&\int_{0}^{r_{0}}\varphi _{0}^{2}\varphi _{1}^{2}rdr.  \label{13}
\end{eqnarray}%
In the range of parameters corresponding to the ultimate vortex confinement, 
$\alpha _{L}$, $\beta _{L}$, $\gamma \sim 0.1-0.5$.\ Let us stress that
these numbers are internally dependent on the external field.

The energy of the island in the vortex-free state $F(L=0)$ can be found from
Eq. (10) by minimization of $F$ with respect to $c_{0}$, while keeping $%
c_{1}=0$:

\begin{equation}
F(L=0)=-\frac{H_{c}(T)^{2}}{\mu _{0}}\xi (T)^{2}d\left( \frac{\alpha _{0}^{2}%
}{4\beta _{0}}\right) .  \label{14}
\end{equation}%
Similarly, the energy of the single-vortex state is given by%
\begin{equation}
F(L=1)=-\frac{H_{c}(T)^{2}}{\mu _{0}}\xi (T)^{2}d\left( \frac{\alpha _{1}^{2}%
}{4\beta _{1}}\right) .  \label{15}
\end{equation}

Now, in order to find a height of the potential barrier for vortex entry, we
start from the vortex-free solution, $c_{1}=0$. We then begin to increase $%
c_{1}$ and we minimize $F$ with respect to $c_{0}$ at fixed $c_{1}$. As a
result, the following "optimal" dependence of $F$ on $c_{1}$ is obtained

\begin{eqnarray}
F(0 &\rightarrow &1)=\frac{H_{c}(T)^{2}}{\mu _{0}}\xi (T)^{2}d\left[ -\frac{%
\alpha _{0}^{2}}{4\beta _{0}}+c_{1}^{2}\left( \alpha _{1}-\frac{\alpha
_{0}\gamma }{\beta _{0}}\right) +c_{1}^{4}\left( \beta _{1}-\frac{\gamma ^{2}%
}{\beta _{0}}\right) \right] ,\text{ \ }c_{1}\leq \sqrt{-\frac{\alpha _{0}}{%
2\gamma },}  \notag \\
F(0 &\rightarrow &1)=\frac{H_{c}(T)^{2}}{\mu _{0}}\xi (T)^{2}d\left(
c_{1}^{2}\alpha _{1}+c_{1}^{4}\beta _{1}\right) ,\text{ \ }\sqrt{-\frac{%
\alpha _{0}}{2\gamma }}<c_{1}\leq \sqrt{-\frac{\alpha _{1}}{2\beta _{1}}},
\label{16}
\end{eqnarray}%
and similarly for the vortex exit, $F(1\rightarrow 0)$. The height of the
potential barrier, which is determined by the saddle point in ($c_{0}$, $%
c_{1}$) space is given by the minimum of $F(0\rightarrow 1)$ with respect to 
$c_{1}$:

\begin{equation}
U_{en}=\frac{H_{c}(T)^{2}}{\mu _{0}}\xi (T)^{2}d\frac{\left( \alpha _{1}-%
\frac{\alpha _{0}\gamma }{\beta _{0}}\right) ^{2}}{4\left( \frac{\gamma ^{2}%
}{\beta _{0}}-\beta _{1}\right) }.  \label{17}
\end{equation}

A systematic study shows that the vortex-free state is unstable with respect
to the vortex penetration at $H\simeq 0.255$ T; while the single-vortex
state becomes unstable with respect to the vortex expulsion at $H\simeq
0.195 $ T. Thus, we see that the maximum hysteresis (in absence of
fluctuations) is $\sim $ 30 \%. Note that the width of the hysteresis region
expands rapidly with increasing $R/\xi (T)$.

\begin{figure}[tbp]
\begin{center}
\includegraphics[width=0.5\textwidth]{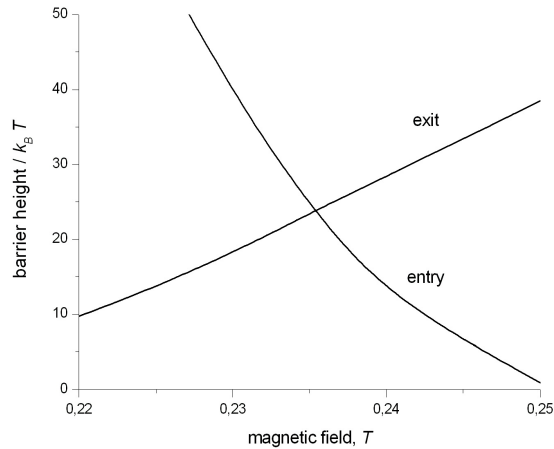}
\end{center}
\caption{ The dependence of the surface barrier height on the
external field $H$.}
\label{Fig1}
\end{figure}

In Fig. 1 we plot the dependence of the barrier height (both for vortex
entry and exit), given in units of $k_{B}T$, on the external field $H$. We
see that heights of both barriers are $\approx $ 23 $k_{B}T$ in the vicinity
of $H=0.235$ T, where energies of the two phases become equal. The ratio of
barrier height and $k_{B}T$ is of the same order as for high-$T_{c}$
materials. The major reasons for a relatively low height of the potential
barrier, as compared to the thermal energy, are (i) very small island
thickness and (ii) very small disc radius which results in the substantial
suppression of the order parameter in the relevant range of applied fields.

\begin{figure}[tbp]
\begin{center}
\includegraphics[width=0.5\textwidth]{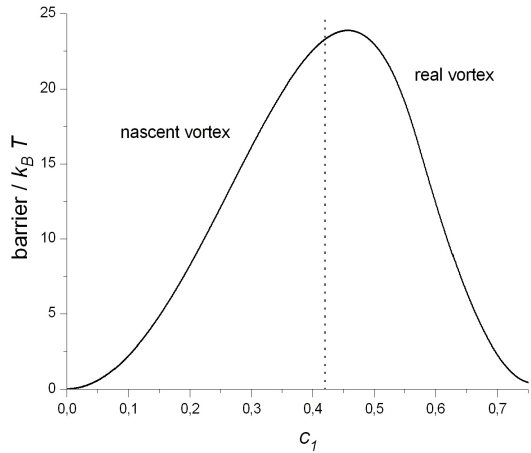}
\end{center}
\caption{ Potential barrier preventing an increase of $c_{1}$, i.e.,
a growth of the population of the Landau level with winding number 1 at $%
H=0.235$ T. This growth describes the formation of the vortex nucleus at the
island edge with the subsequent motion of a formed vortex towards the island
center. Dotted line separates the state with the "nascent" vortex, when the
order parameter is still nonzero everywhere, and the phase with the real
vortex.}
\label{Fig2}
\end{figure}

In Fig. 2 we show the calculated surface barrier for the vortex entry, given
by Eqs. (16), as a function of $c_{1}$. It is quite remarkable that almost
the whole interval of $c_{1}$ from $c_{1}=0$ up to the saddle point
corresponds to the vortex core nucleation rather than to the motion of a
formed vortex inside the island. Fig. 2 thus demonstrates that, under the
experimental conditions of Ref. \cite{Cren}, a vortex position cannot
be an appropriate variable to analyze thermal activation over the surface
barrier, due to very small island's lateral dimensions.

\section{Viscosity}

When vortices move, local magnetic field inside a superconductor changes in
time. According to the Faraday law, this change produces a nonzero electric
field which leads to an energy dissipation. The dissipation can be
interpreted in terms of a viscous force acting on moving vortices.

The microscopic theory of this phenomenon is rather complicated. However,
reasonable estimates for the viscosity force can be obtained by using
phenomenological approach within the Ginzburg-Landau theory \cite%
{Tinkham,UFN,Shmidt,AL}. In this paper, in order to estimate viscosity
associated with the \textit{motion of the order parameter}, we apply similar
ideas.

When finding the island energy, we did not take into account that the
magnetic field varies inside the sample. This is fully justified since the
disc is very thin ($d\ll \lambda (T)$) and its radius is also small, $R\sim
\lambda (T)$, that allows one to use a perturbation theory%
\cite{Chapman} in terms of a small parameter $1/\kappa _{eff}^{2}$,
where $\kappa _{eff}$ is much larger than $\kappa $ due to demagnetization
effects. Physically, this implies that the additional field generated by the
Meissner current and by the vortex is small in comparison with the external
uniform field. If we know the solution of the first Ginzburg-Landau equation 
$f$ for the order parameter in leading order, the vector potential $\mathbf{a%
}_{add}$ of this additional field can be found as%
\begin{equation}
\mathbf{a}_{add}\mathbf{=}\frac{1}{4\pi }\int \frac{\mathbf{j}}{r}dV,
\label{18}
\end{equation}%
where an integration is performed over the island volume and $\mathbf{j}$ is
a supercurrent density given by%
\begin{equation}
\mathbf{j=-}\frac{1}{\kappa ^{2}}\left[ \mathbf{a}|f|^{2}+\frac{i}{2}\left(
f^{\ast }\mathbf{\nabla }f-f^{\ast }\mathbf{\nabla }f\right) \right] ,
\label{19}
\end{equation}%
where $\mathbf{a}=\mathbf{e}_{\varphi }h_{e}r/2$ is the dimensionless vector
potential in leading order which corresponds to the applied uniform field $%
h_{e}$. Instead of using the exact solution for $f$, we take the lowest
Landau level approximation, as done in Section II.

Vortex entry or exit is associated with $f$ changing in time. This change
generates the local electric field $\mathbf{e}$, which can be found from the
Faraday law%
\begin{equation}
\text{rot}\ \mathbf{e=-}\frac{\partial \mathbf{(}\text{rot }\mathbf{a}_{add}%
\mathbf{)}}{\partial t},  \label{20}
\end{equation}%
and hence%
\begin{equation}
\mathbf{e=-}\frac{\partial \mathbf{a}_{add}}{\partial t},  \label{21}
\end{equation}%
where $\mathbf{a}_{add}$ is given by Eq. (18).

The order parameter inside the island is already strongly suppressed in the
vicinity of a transition to the one-vortex state. Therefore, the
conductivity is close to the conductivity of Pb in the normal state $\sigma
_{n}$. Then, the dissipation rate in the sample $W$ can be estimated as 
\begin{equation}
W/\left( H_{c2}(T)^{2}\xi (T)^{5}\right) \mathbf{\sim }\sigma _{n}\int 
\mathbf{e}^{2}dV,  \label{22}
\end{equation}%
where the denominator in the left-hand side comes from the normalization
condition.

Let us now consider the process of vortex entry within our representation
for the order parameter, as given by Eq. (6). Vortex penetration is
associated with the growth in time of $c_{1}$ \textit{starting} from $0$,
along the optimal path. It is then easy to see, from Eq. (19), that the 
\textit{leading-order} in $c_{1}$ contribution to $\mathbf{j}$ has the
following components%
\begin{eqnarray}
j_{\varphi } &=&-\frac{1}{\kappa ^{2}}c_{0}c_{1}f_{1}f_{0}\left( h_{e}r-%
\frac{1}{r}\right) \cos \varphi \text{,}  \notag \\
j_{r} &=&-\frac{1}{\kappa ^{2}}c_{0}c_{1}\left( f_{1r}^{\prime
}f_{0}-f_{0r}^{\prime }f_{1}\right) \sin \varphi \text{.}  \label{23}
\end{eqnarray}%
We substitute Eqs. (23) to Eq. (18). Since we are interested in $\mathbf{a}%
_{add}$ \textit{inside} a thin disc, $d\ll R$, we can neglect the dependence
on $z$ in the denominator of the integrand in the right-hand side (RHS) of
Eq. (18). Then, for the radial component $\mathbf{a}_{add}^{(r)}$, after
performing integration on angle, we obtain%
\begin{equation}
\mathbf{a}_{add}^{(r)}\mathbf{=}\frac{d\cos \varphi }{2\pi \xi (T)\kappa ^{2}%
}c_{0}c_{1}g(r),  \label{24}
\end{equation}%
where

\begin{equation}
g(r)\mathbf{=}\frac{1}{r}\int_{0}^{r_{0}}r_{1}^{-1}\left( f_{1r_{1}}^{\prime
}f_{0}-f_{0r_{1}}^{\prime }f_{1}\right) \left[ -(r_{1}+r)E\left( \frac{2%
\sqrt{rr_{1}}}{r+r_{1}}\right) +\frac{r^{2}+r_{1}^{2}}{r+r_{1}}K\left( \frac{%
2\sqrt{rr_{1}}}{r+r_{1}}\right) \right] dr_{1},  \label{25}
\end{equation}%
where $K$ and $E$ are complete elliptic integrals of the first and second
kind, respectively. Similarly, $\mathbf{a}_{add}^{(\varphi )}$ is found,
which is proportional to $\sin \varphi $, while the integrand in the RHS of
Eq. (25) now contains $f_{0}f_{1}\left( h_{e}r_{1}-1/r_{1}\right) $ instead
of $\left( f_{1r_{1}}^{\prime }f_{0}-f_{0r_{1}}^{\prime }f_{1}\right) $. The
motivation to calculate viscosity only in leading order in $c_{1}$\ will be
clarified in Section IV.

By using Eq. (21), we find the local electric field $\mathbf{e}$ as a
function of the \textit{time derivative} $c_{1t}^{\prime }$%
\begin{equation}
e_{r}=-c_{1t}^{\prime }\cos \varphi \frac{d}{2\pi \xi (T)\kappa ^{2}}\sqrt{-%
\frac{\alpha _{0}}{2\beta _{0}}}g(r),  \label{26}
\end{equation}%
and similarly for $e_{\varphi }$. The dissipation rate is then determined
from Eq. (22) as%
\begin{equation}
W\mathbf{\sim }\left( c_{1t}^{\prime }\right) ^{2}\sigma _{n}\delta \frac{%
H_{c}(T)^{2}\xi (T)^{2}d^{3}}{\kappa ^{2}},  \label{27}
\end{equation}%
where $\delta $ is found from integrating numerically the $r$-dependence of $%
\mathbf{e}$, as given by Eq. (25), and by using calculated values of $\alpha
_{0}$\ and $\beta _{0}$. We then arrive to the simple estimate%
\begin{equation}
\delta \sim 10^{-2}.  \label{28}
\end{equation}

The "force" of resistance against the growth of $c_{1}$ is given by $W$
divided by $c_{1t}^{\prime }$. This "force" is proportional to $%
c_{1t}^{\prime }$. The viscosity coefficient $\eta ^{(c_{1})}$, i.e., the
proportionality factor, reads%
\begin{equation}
\eta ^{(c_{1})}\mathbf{\sim }\sigma _{n}\frac{H_{c}(T)^{2}\xi (T)^{2}d^{3}}{%
\kappa ^{2}}10^{-2}.  \label{29}
\end{equation}%
The same expression is found for the viscosity coefficient associated with
the growth of $c_{0}$ (the\ case of a vortex exit). The estimate for $\delta 
$\ remains valid as well.

We stress that $\eta ^{(c_{1})}$, by definition, is directly linked to the
change of the population of the level with winding number 1 and not to the
change of the vortex position.

From Eq. (29) we see that $\eta ^{(c_{1})}$ is proportional to $d^{3}$ and
not to $d$, as can be naively expected. Such a strong dependence on $d$ is
due to demagnetization effects that significantly diminish the diamagnetic
response of the sample and, hence, the electric field. This response is
further suppressed due to extremely small lateral dimensions of the island,
because the supercurrent created by the vortex is localized within a
confined area $\sim \xi (T)^{2}$.

\section{First passage time}

In Sections II and III we have found a profile of the potential barrier and
the viscosity in terms of $c_{1}$ ($c_{0}$). We now are in a position to
find the typical time for the vortex penetration (expulsion). This can be
done by using the backward Fokker-Planck equation, also written in terms of $%
c_{1}$ ($c_{0}$).

Let us first consider vortex entrance. The dependence $U(c_{1})$ of the
system energy on $c_{1}$\ is plotted in Fig. 2. Initially, the system is in
the left-hand potential well (vortex-free state, $c_{1}=0$). To calculate
the average time for the transition to the right-hand potential well
(one-vortex state) we use the formalism developed in Ref. \cite{FP}.

We introduce the probability $p(c_{1}^{\prime },t\mid c_{1},0)$ to find the
system at $c_{1}^{\prime }$ in the moment of time $t$, \textit{provided} it
was at $c_{1}$ in the moment $t=0$. The relevant interval of possible values
for $c_{1}$ is between the two potential wells, as shown in Fig. 2 ($%
c_{1}\in \left[ a,b\right] $, where $a=0$, $b=\sqrt{-\frac{\alpha _{1}}{%
2\beta _{1}}}$). We also denote the saddle-point value of $c_{1}$ as $s$.
Then, the probability $G(c_{1},t)$ that the system is still in this interval
is%
\begin{equation}
G(c_{1},t)\mathbf{=}\int_{a}^{b}p(c_{1}^{\prime },t\mid
c_{1},0)dc_{1}^{\prime }.  \label{30}
\end{equation}

Since the system is homogeneous in time, we can replace $p(c_{1}^{\prime
},t\mid c_{1},0)$ by $p(c_{1}^{\prime },0\mid c_{1},-t)$, for which the
backward Fokker-Planck equation applies 
\begin{equation}
\frac{\partial p(c_{1}^{\prime },0\mid c_{1},-t)}{\partial (-t)}\mathbf{=}%
\frac{1}{\eta ^{(c_{1})}}\frac{\partial U}{\partial c_{1}}\frac{\partial
p(c_{1}^{\prime },0\mid c_{1},-t)}{\partial c_{1}}-\frac{k_{B}T}{2\eta
^{(c_{1})}}\frac{\partial ^{2}p(c_{1}^{\prime },0\mid c_{1},-t)}{\partial
c_{1}^{2}}.  \label{31}
\end{equation}%
We can now switch back to $p(c_{1}^{\prime },t\mid c_{1},0)$ and to
integrate both sides of Eq. (31) on $c_{1}\in \left[ a,b\right] $. Taking
into account Eq. (30), we obtain an equation for $G(c_{1},t)$:%
\begin{equation}
\frac{\partial G(c_{1},t)}{\partial t}\mathbf{=-}\frac{1}{\eta ^{(c_{1})}}%
\frac{\partial U}{\partial c_{1}}\frac{\partial G(c_{1},t)}{\partial c_{1}}+%
\frac{k_{B}T}{2\eta ^{(c_{1})}}\frac{\partial ^{2}G(c_{1},t)}{\partial
c_{1}^{2}}.  \label{32}
\end{equation}

We also impose a condition%
\begin{equation}
p(c_{1}^{\prime },0\mid c_{1},0)\mathbf{=\delta (}c_{1}-c_{1}^{\prime }%
\mathbf{)},  \label{33}
\end{equation}%
which shows that at $t=0$ the system was in the definite state. From this
condition and Eq. (30), we get%
\begin{equation}
G(c_{1},0)\mathbf{=}1,  \label{34}
\end{equation}%
which evidences that the system was inside the interval $c_{1}\in \left[ a,b%
\right] $ at $t=0$.

We then assume a reflecting condition at $c_{1}=a$, suggesting that the
system never leaves $\left[ a,b\right] $ interval through this end. For the
opposite limit, $c_{1}=b$, we impose an absorbing condition, which means
that once the system reaches this point, it never enters again the $\left[
a,b\right] $ interval. We now calculate the average time $\tau (c_{1})$,
which corresponds to the time, system spends in the interval, \textit{%
provided} it was at some fixed $c_{1}$ in the initial moment. Thus, $\tau
(a) $\ will give us the first passage time from the well at $c_{1}=a$
(vortex-free state) to the well at $c_{1}=b$ (one-vortex state). The time $%
\tau (c_{1})$ can be expressed from $G(c_{1},t)$. We first note that $%
G(c_{1},t)$ gives the probability that the system at the moment $t$ is still
within the interval $\left[ a,b\right] $, while $G(c_{1},t+dt)$ is the
probability that it is still there at $t+dt$. The difference, $-\frac{%
\partial G(c_{1},t)}{\partial t}dt$, represents the probability for the exit
within the time interval $\left[ t,t+dt\right] $. Then, the average exit
time $\tau (c_{1})$ is given by%
\begin{equation}
\tau (c_{1})\mathbf{=}\mathbf{-}\int_{0}^{\infty }dt\left\{ \frac{\partial
G(c_{1},t)}{\partial t}\right\} =-\int_{0}^{\infty }dG(c_{1},t).  \label{35}
\end{equation}

Using Eq. (32) for $G(c_{1},t)$, we can obtain the differential equation for 
$\tau (c_{1})$ by integrating both sides of Eq. (32) on $t$ from 0 to $%
\infty $:%
\begin{equation}
\mathbf{-}\frac{1}{\eta ^{(c_{1})}}\frac{\partial U}{\partial c_{1}}\frac{%
\partial \tau (c_{1})}{\partial c_{1}}+\frac{k_{B}T}{2\eta ^{(c_{1})}}\frac{%
\partial ^{2}\tau (c_{1})}{\partial c_{1}^{2}}=-1,  \label{36}
\end{equation}%
where we have taken into account Eq. (34). Eq. (36) defines a first passage
time from arbitrary $c_{1}\in \left[ a,b\right] $ to $b$. This equation must
be supplemented by the conditions for the reflecting boundary at $a$ and
absorbing boundary at $b$

\begin{eqnarray}
\left. \frac{\partial \tau (c_{1})}{\partial c_{1}}\right\vert _{c_{1}=a}
&=&0,  \label{37} \\
\tau (b) &=&0.  \label{38}
\end{eqnarray}%
The reflecting condition can be obtained from the vanishing of the
probability current, while absorbing condition means that the system
immediately exits the interval, once it is initially put at the absorbing
edge.

Ordinary differential equation (36) with boundary conditions (37) and (38)
can be solved by standard methods. Finally, for $\tau (a)$, which will be
below denoted as $\tau $, we obtain

\begin{equation}
\tau \equiv \tau (a)\mathbf{=}\frac{2}{k_{B}T}\int_{a}^{b}\left\{ \exp
\left( \frac{U(y)}{k_{B}T}\right) \int_{a}^{y}\eta ^{(c_{1})}\exp \left( 
\frac{-U(x)}{k_{B}T}\right) dx\right\} dy.  \label{39}
\end{equation}%
When integrating over $y$, the dominant contribution to the RHS of Eq. (39)
is given by those values of $y$ for which $\exp \left( U(y)/k_{B}T\right) $
is largest, i.e., by the neighborhood of the saddle point $s$ (see Fig. 2).
At the same time, the internal integral is essentially constant in this
area. It is therefore can be approximated by replacing $y$ in the upper
limit of integration by $b$. The RHS of Eq. (39) is then rewritten through a
product of two independent integrals%
\begin{equation}
\tau \mathbf{\simeq }\frac{2}{k_{B}T}\left( \int_{a}^{b}\exp \left( \frac{%
U(y)}{k_{B}T}\right) dy\right) \left( \int_{a}^{b}\eta ^{(c_{1})}\exp \left( 
\frac{-U(x)}{k_{B}T}\right) dx\right) .  \label{40}
\end{equation}%
The very high accuracy of such a transformation can be easily verified
numerically. It is directly linked to the fact that the barrier height is
large compared to $k_{B}T$.

The dominant contribution to the first integral in the RHS of Eq. (40) is
provided by the neighborhood of $s$, while for the second integral it is
given by the neighborhood of $a$. The latter is basically the reason why, in
Section III, we have calculated the viscosity coefficient $\eta ^{(c_{1})}$
only in leading order in $c_{1}$: taking into account higher-order terms
does not lead to any significant changes. We now expand $U$ in the vicinity
of $s$ and $a$ as%
\begin{equation}
U(x\approx s)\mathbf{\approx }U(s)-\frac{\left\vert U^{\prime \prime
}(s)\right\vert }{2}(x-s)^{2},  \label{41}
\end{equation}%
\begin{equation}
U(x\approx a)\mathbf{\approx }U(a)+\frac{\left\vert U^{\prime \prime
}(a)\right\vert }{2}(x-a)^{2}.  \label{42}
\end{equation}%
It is easy to see that for the simple potential, given by Eq. (16), $%
\left\vert U^{\prime \prime }(s)\right\vert =2\left\vert U^{\prime \prime
}(a)\right\vert $. After substituting Eqs. (41) and (42) to Eq. (40) and
integrating, we obtain%
\begin{equation}
\tau \mathbf{=}\tau _{0}\exp \left( \frac{U_{en}}{k_{B}T}\right) ,
\label{43}
\end{equation}%
where $U_{en}=U(s)-U(a)$ is the barrier height, as given by Eq. (17). The
"attempt time" $\tau _{0}$ is 
\begin{equation}
\tau _{0}\mathbf{\approx }\frac{\pi \sqrt{2}\eta ^{(c_{1})}}{U^{\prime
\prime }(a)}.  \label{44}
\end{equation}%
By using Eq. (16), for $U^{\prime \prime }(a)$ we get%
\begin{equation}
U^{\prime \prime }(a)=\frac{2H_{c}(T)^{2}}{\mu _{0}}\xi (T)^{2}d\left(
\alpha _{1}-\frac{\alpha _{0}\gamma }{\beta _{0}}\right) ,  \label{45}
\end{equation}%
with $\left( \alpha _{1}-\alpha _{0}\gamma /\beta _{0}\right) \sim 0.1$.

Thus, Eqs. (44) and (45) together with Eq. (29) for $\eta ^{(c_{1})}$ yield
the following estimate for $\tau _{0}$%
\begin{equation}
\tau _{0}\mathbf{\sim }\sigma _{n}\mu _{0}\frac{d^{2}}{\kappa ^{2}}.
\label{46}
\end{equation}%
We may assume that the conductivity is linear in $T$, so from the known $%
\sigma _{n}$ of Pb at room temperature we can estimate $\sigma _{n}$ at the
conditions of the experiment\cite{Cren} as $\sigma _{n}$ $\gtrsim
10^{8}$ $\Omega ^{-1}m^{-1}$. Then, $\tau _{0}$ is $\sim $10$^{-15}-$10$%
^{-14}$s. This value together with our result for the surface barrier
height, $U_{en}\approx 23k_{B}T$, gives $\tau \sim 10^{-5}-$10$^{-4}$ s. The
very small value for $\tau _{0}$ is due to the fact that the supercurrent is
circulating within the tiny volume $\sim \xi (T)^{2}d$, so that the magnetic
response of the island is quite weak. Moreover, the order parameter is
significantly suppressed. In addition, the distance from the island's edge
to its center, which vortex has to pass, is very short.

The above derivation was presented for the case of a vortex entry. The
evaluation of the typical time for the vortex exit is quite similar. We have
found that Eq. (46) remains applicable, which leads to similar
order-of-magnitude estimates for $\tau $.

Experimentally, it has also been observed that the magnetic field
corresponding to the transition between one-vortex and vortex-free states
was reversible within few percents\cite{Cren}. This can be understood
by noting that, when changing the magnetic field away from 0.235 T, barriers
for exit and entrance become asymmetrical. This means that the equilibrium
probability distribution becomes extremely strongly peaked within the deeper
well, as follows directly from Fig. 2. Indeed, it is easy to see that the
difference in \textit{heights} of the \textit{barriers} gives the difference
in \textit{energies} between the \textit{bottoms} of the two wells, because $%
\left( U(s)-U(a)\right) -\left( U(s)-U(b)\right) $ $=U(b)-U(a)$. Since the
absolute value of the difference in heights far exceeds $k_{B}T$ already in
the vicinity of 0.235 T (as seen from Fig. 2), the probability to find a
system in a more shallow well is $\sim \exp (-\left\vert
U(a)-U(b)\right\vert /k_{B}T)<<1$. Thus, the system gets stabilized in the
deeper well.

Note that $\tau _{0}$, as given by Eq. (46), is not explicitly dependent on $%
T$, in contrast to the similar quantity calculated in Ref. %
\cite{PetChech} within the London approximation. Technically, this is
due to the fact that the barrier profile, as given by Eq. (16), is
proportional to the \textit{second} power of the relevant variable (with
which $\eta $ is associated), i.e., to $c_{1}^{2}$ at $c_{1}\rightarrow 0$,
while the barrier profile in Ref. \cite{PetChech} is proportional to
the \textit{first} power of vortex displacement when this displacement is
small. Through the expansion (42), this affects the second integral in the
RHS of Eq. (40) and therefore leads to the mentioned difference. Physically,
it might be attributed to the fact that vortex penetration processes for the
cases of a half-infinite superconductor and the island in the ultimate
vortex confinement regime are qualitatively different. Indeed, in the first
case this process is associated with the motion of a "rigid" vortex line
inside the superconductor, while in the second case it is represented by the
formation of a "soft" vortex nucleus at the edge of the disc. In the first
case, the barrier is due to the pronounced magnetic interaction with the
superconductor's surface. In the second case, it is due to the condensation
energy increase and the residual magnetic interactions.

We also wish to notice that the effect of thermal fluctuations of the order
parameter on vortex penetration and expulsion in mesoscopic superconductors
was previously studied\cite{Hern} by Hernandes \textit{et al.} by the
numerical integration of the time-dependent Ginzburg-Landau equations. In
these computations no thermal activation of vortices over the surface
barrier was observed for parameters corresponding to samples made of low-$%
T_{c}$ superconductors, while this activation was shown to be possible for
those samples, which are made of high-$T_{c}$ materials. This is consistent
with previous experiments on Al discs\cite{Geim} and micron-size
structures made of high-$T_{c}$ material\cite{Kirtley}. We think that
the difference between the results of the present paper and of Ref. %
\cite{Hern} concerning low-$T_{c}$ samples can be attributed to the
fact that in calculations of Ref. \cite{Hern} the main parameters were
chosen in such a way as to model experiments\cite{Geim}, where (i)
discs of much larger thickness were studied compared to the island of Ref. %
\cite{Cren} (one order of magnitude and more), (ii) the order
parameter was not significantly suppressed due to larger discs radii, (iii)
sample temperature was lower in one order of magnitude or so compared to
that of Ref. \cite{Cren} ($T_{c}$ of Al is at least several times
lower than $T_{c}$ of Pb).

In the very recent experiment\cite{Bartolf}, strong \textit{%
current-induced} fluctuation phenomena were observed in ultrathin ($\approx
6 $ nm) superconducting nanowires fabricated in a form of a meander and made
of NbN. Lateral dimensions of sample's typical segment, in terms of $\xi (T)$%
, were one order of magnitude larger than that for the island in the
ultimate vortex confinement regime. Structures of this kind are used in
single-photon detectors, while fluctuations are major source of dark-count
events in such devices. Fluctuation phenomena were explained by a
thermally-activated vortex entry, as well as by the unbinding of
vortex-antivortex pairs. The height of the surface barrier, as calculated in
Ref. \cite{Bartolf}, was at least one order of magnitude larger (in
terms of $k_{B}T$) than the one for the superconducting island studied in
the present paper. Nevertheless, vortex penetration events were detectable.

Ref. \cite{Nishio} deals with the experimental investigation of
superconducting islands of Pb, which were quite similar to the islands
studied in Ref. \cite{Cren} and in the present paper. The thickness of
those islands was even smaller than that of Ref. \cite{Cren}. However,
measurements were performed at significantly lower temperature of 2.0 K.
Probably, this is why some hysteresis for vortex penetration and expulsion
was observed. Indeed, the activation time depends \textit{exponentially} on
the height of the potential barrier \textit{this height being much larger}
than $k_{B}T$. Nevertheless, it was mentioned in Ref. \cite{Nishio}
that the width of the hysteresis region was noticably smaller than the
theoretical values obtained from the Ginzburg-Landau theory. The results of
the present paper evidence that thermal fluctuations might be responsible
for an observed shrinkage of the hysteresis region.

Let us finally mention a very recent experiment\cite{Pekin} on \textit{%
one-atomic layer} Pb films, where magnetic vortices were directly
visualized. Thermal fluctuations must be strongest in such ultimately thin
films.

Since thermal fluctuations are more pronounced in low-dimensional systems,
an understanding of vortex matter fluctuations in thin and small
superconducting structures is of certain importance. It is howevere clear
that more intensive experimental, as well as theoretical efforts are needed
to reveal systematically those confinement and temperature regimes which
favor thermal fluctuations in such nanostructures or, on the contrary, which
enable one to safely avoid them.

\section{Conclusions}

Although it is generally believed that the surface barrier in \textit{low}-$%
T_{c}$ superconductors is too high to be suppressed by thermal fluctuations,
in this paper we have shown that such a scenario is possible in small and
ultrathin superconducting samples studied in modern experiments. In
particular, we have demonstrated that an absence of hysteresis for vortex
entry and exit in the nanosized island of Pb observed in the very recent
experiment of Ref. \cite{Cren} (and the shrinkage of hysteresis region
reported in the similar experiment of Ref. \cite{Nishio}) can be
explained in terms of a thermal activation of a vortex over the surface
barrier.

Lateral dimensions of these islands were so small that they could
accommodate only one vortex before gradual transition to the normal state.
For this reason, London approximation, which was previously applied to
describe thermal surmounting of a surface barrier by vortices, is not
applicable.

We therefore have developed an alternative theoretical approach based on the
lowest Landau level approximation for the Ginzburg-Landau order parameter,
which is suitable for such extremely small structures. The surface barrier
profile was calculated in terms of populations of the two relevant Landau
levels being used throughout the paper as "good" variables instead of the
"bad" one (vortex position). The viscosity coefficient, associated with the
motion of the projections of the order parameter, was estimated with account
of demagnetization effects. Finally, the expression for the typical time of
thermally-assisted vortex entry (exit) was obtained from the Fokker-Planck
equation, also written in terms of the populations of Landau levels. This
expression is of Arrhenius type.

We have found that the barrier height, under the conditions of experiment %
\cite{Cren}, is nearly 20 in terms of a thermal energy $k_{B}T$, which
is close to the similar quantity for high-$T_{c}$ materials, where thermal
suppression of a surface barrier is observable. Although the exponent of
this ratio is still very large, the preexponential "attempt time" is quite
small, so that the typical time both for the vortex exit and entrance is
several orders of magnitude smaller than 1 s.

\section{Acknowledgements}

Useful discussions with D. Roditchev, T. Cren, and A. L. Rakhmanov are
acknowledged. This work was supported by RFBR (project no. 09-02-00248) and
"Dynasty Foundation".

\end{document}